\documentclass[twocolumn,showpacs,preprintnumbers,amsmath,amssymb,aps]{revtex4}
\usepackage{graphicx}
\usepackage{epsfig}
\def\lsim{\raise0.3ex\hbox{$\;<$\kern-0.75em\raise-1.1ex\hbox{$\sim\;$}}}
\def\gsim{\raise0.3ex\hbox{$\;>$\kern-0.75em\raise-1.1ex\hbox{$\sim\;$}}}
\newcommand{\be}{\begin{eqnarray}}
\newcommand{\ee}{\end{eqnarray}}

\def\bea{\begin{eqnarray}}
\def\eea{\end{eqnarray}}

\begin{document}
\title{Right-handed Sneutrino Inflation in SUSY $B-L$ with Inverse Seesaw}
\author{Shaaban Khalil$^{1,2}$, and Arunansu Sil$^{3}$  }
\vspace*{0.2cm}
\affiliation{$^1$Center for Theoretical Physics at the
British University in Egypt, Sherouk City, Cairo 11837, Egypt.\\
$^2$Department of Mathematics, Faculty of
Science,  Ain Shams University, Cairo, 11566, Egypt.\\
$^3$Department of Physics, Indian Institute of Technology, Guwahati,
Assam 781039, India.}

\begin{abstract}
\noindent We propose a scenario for realizing inflation in the
framework of supersymmetric $B-L$ extension of the
Standard Model. We find that one of the associated right-handed sneutrinos 
(the superpartner of the Standard Model singlet fermion) can provide a new 
non-trivial inflationary trajectory at tree level (therefore breaking $B-L$ during 
inflation). As soon as the inflation ends, the right-handed sneutrino falls into the 
supersymmetric vacuum, with a vanishing vacuum expectation value, so that $B-L$ 
symmetry is restored. In this class of models, the $B-L$ gauge symmetry will be 
radiatively broken at a TeV scale and light neutrino masses are generated through 
the inverse seesaw mechanism. 
\end{abstract}
{\bf \pacs{98.80.Cq, 14.60.St, 11.30.Pb}}
\maketitle
%

Inflation \cite{Guth:1980zm} is one of the most interesting scenarios for explaining
the origin of structure formation in the Universe.
It leads to a very consistent result with the recent observations on 
the cosmic microwave background radiation \cite{Komatsu:2010fb}. 
The basic idea of inflation is based on a scalar field, called inflaton 
dominating the energy density of the universe, rolling down slowly 
along an almost flat potential at some early stage of the universe. 
As a consequence, an extremely rapid exponential expansion of the 
universe took place. After inflation ends, the field stabilizes in the 
global (true) minimum of the potential and the universe enters into the 
radiation dominated era through reheating. Though it is very intriguing 
to find an inflationary potential from a particle physics model, at the same
time this sort of picture is difficult to construct also. The
difficulty arises mostly in realizing the flatness of the
potential. Several attempts have been made in recent
years \cite{review}. There is presently no agreed upon model that
can describe inflation from particle physics, however it is now
established that the construction requires physics beyond the 
Standard Model (SM) of particle physics.

An extension of the SM gauge group seems also mandatory for
accommodating the non-vanishing neutrino masses confirmed by
neutrino oscillation experiments \cite{Wendell:2010md}. Recently,
it has been shown that the simplest possible such an extension of
the SM is the one based on the gauge group $SU(3)_C \times SU(2)_L
\times U(1)_Y \times U(1)_{B-L}$~\cite{Khalil:2006yi}. In this
class of model, it turns out that the scale of $B-L$ symmetry
breaking can be related to the scale of supersymmetry (SUSY)
breaking \cite{Khalil:2007dr}. Therefore, a TeV scale type I or
inverse seesaw mechanism can be naturally implemented
\cite{Khalil:2010zz}. In recent years there has been considerable
interest in analyzing the phenomenological implications of this
class of models and their possible signatures at the Large Hadron
Collider (LHC) \cite{lhc}.

In this article we consider the SUSY $B-L$ extension of the SM
with inverse seesaw, where right-handed sneutrino plays the role
of inflaton. We show that a new non-trivial inflationary
trajectory can be realized at tree level, along which the $B-L$
gauge symmetry is broken (during inflation), through a large field
value of right-handed sneutrino. At the end of inflation, the
right-handed sneutrino continues falling into the SUSY vacuum, with
a vanishing sneutrino vacuum expectation value (vev). Therefore,
$B-L$ symmetry is restored which is important since in the
scenario we are interested in (the SUSY $B-L$ extension
\cite{Khalil:2010zz}), the $B-L$ breaking scale is $\sim$ TeV. In
our set-up, there is one more field involved in the superpotential
for inflation apart from the right handed sneutrino, a singlet under 
the entire gauge group ($X$), whose
role is to put the inflaton into the right direction during
inflation and most importantly to nullify any possible
contribution to the neutrino mass matrix by getting a vev at the
end of inflation. Our effective inflationary potential then turns
out to be similar with the one in chaotic inflation \cite{Linde:1983gd}  
though it is different to start with. Furthermore the dynamics is more involved
as we will see later.

Chaotic inflation is a kind of simplest possibility to realize inflation. It can be incorporated
in a supersymmetric framework through sneutrino inflation
\cite{Murayama:1992ua}. Here we discuss in brief the basic
structure of it which incorporates minimal supersymmetric seesaw
model for neutrinos in contrast with our set-up with inverse
seesaw mechanism. It is usually assumed that right-handed neutrino
superfied, $N$, is odd under a $Z_2$ symmetry in order to forbid
its cubic term in the superpotential. Therefore the sneutrino
effective potential is given by $V = \frac{1}{2} M^2_N \vert
\tilde{N}\vert^2$, where $M_N$ is the mass of right-handed
neutrino. Such a scenario leads to chaotic inflation, where the
universe starts with a random value (chaotic) of the sneutrino 
inflaton $\tilde{N}$. It can be noted that in this case the $M_N$ is
fixed to be $\sim 10^{13}$ GeV since the amplitude of primordial
density perturbation is directly proportional to $M_N$. In the 
framework of inverse seesaw, no elements of the neutrino mass
matrix is expected to be that large. So we do not expect the above
type of sneutrino inflation to take place with inverse seesaw.
We have introduced the $X$ field, interplay of which
with the right handed neutrino superfield can provide chaotic
inflation and simultaneously ensures that there is no such large
contributions present in the neutrino mass matrix, thereby the
inverse seesaw structure remains intact. With this modification,
it also fits the cosmological data. The possibility of hybrid inflation with
sneutrino inflaton and with non-minimal K\"ahler potential has
been considered in Ref.\cite{Antusch:2004hd}. Another class of 
models where the inflationary direction involves the right handed 
sneutrino field can be found in \cite{other-sneutrino-inf}.

Before going to discuss the inflationary picture, let us discuss
in brief the construction of the SUSY $B-L$ extension of the SM
with inverse seesaw mechanism. The particle content of this model
includes the following superfields in addition to those in the
MSSM: two SM singlet chiral Higgs superfields $\chi_{1,2}$ (the
vev of their scalar components spontaneously break the
$U(1)_{B-L}$ at the TeV scale and $\chi_2$ is required to cancel
the $U(1)_{B-L}$ anomaly); three sets of SM singlet chiral
superfields $N_i, S_{1_i}, S_{2_i} (i =1,2,3)$ (to implement the
inverse see-saw mechanism and also to cancel $B-L$ anomaly). The
superpotential of the leptonic sector in this model is given by
\cite{Khalil:2010zz}
{\small \begin{equation}%
W\!=\!Y_ELH_1E^c\!+\!Y_{\nu}LH_2N^c\!+\!Y_SN^c\chi_1S_2\!+\!\mu
H_1H_2\!+\!\mu'\chi_1\chi_2.%
\end{equation}}%
The charge specification of superfields, appeared in this
superpotential, under $U(1)_{B-L}$ and $U(1)_R$ are given in Table
I.
\begin{table}[h]
\begin{center}
\begin{tabular}{|c|c|c|c|c|c|c|c|c|c|}
\hline ${\rm {Charges}} $ & $E^c$ & $L$ & $H_1$ & $H_2$ & $N^c$ &
$\chi_1$ &$\chi_2$
& $S_1$ & $S_2$ \\
\hline $B-L$ & $1$ & $-1$ & $0$ & $0$ & $1$ & $1$ &$-1$
& $2$ & $-2$ \\
\hline $R$ & $2$ & $0$ & $0$ & $2$ & $0$ & $1$ &$1$
& $1$ & $1$ \\
\hline
\end{tabular}
\end{center}
\caption{\small Charges under $U(1)_{B-L}$ and $U(1)_R$ .}
\end{table}
A term proportional to $S_1S_2$ is also allowed by the symmetry, which we will discuss later in the
context of inflationary superpotential along with the term involved with $X$ field.

As mentioned above, the $B-L$ symmetry is radiatively broken by
the non-vanishing VEVs $\langle \chi_1 \rangle = v^{\prime}_1$ and
$\langle \chi_2 \rangle = v^\prime_2$ \cite{Khalil:2007dr}. After
$B-L$ and EW symmetry breaking, the neutrino Yukawa
interaction terms lead to the following expression:
\be
{\cal L}_m^{\nu} = m_D \bar{\nu}_L N^c + M_N {\bar {N^c}} S_2 + 
{\rm {~h.c.}},
\ee
where $m_D=Y_{\nu} v\sin\beta$, $ M_N = Y_{S} v' \sin\theta$
and $v'^2 = v'^2_1 + v'^2_2 $.
In this framework, the light neutrino masses are related to a
small mass term $\mu_s S^2_2$ in the Lagrangian, with $\mu_s\sim {\cal O}(1)$ {\rm
KeV} which can be generated at $B-L$ scale through
non-renormalizable higher order term $\frac{\chi_1^4 S^2_2}{M_I^3}$, with $M_I$ is an intermediate scale of order ${\cal
O}(10^7)$ GeV. Therefore, the following light and heavy neutrino
masses are
obtained respectively:%
\bea %
m_{\nu_{\ell}}&=& \frac{m_D^2\mu_{s}}{M_N^2+m_D^2},\\
m_{\nu_{H,H'}}&=&\pm \sqrt{M_N^2+m_D^2}+\frac{1}{2}\frac{M_N^2
\mu_{s}}{M_N^2+m_D^2}.%
\eea A noticeable difference with the usual seesaw mechanism is
that in this case, no terms in the neutrino mass matrix need to be
very large.

In this type of models, the sneutrino is given by linear
combination of $(\tilde\nu_L, \tilde\nu^{\dagger}_L, \tilde N,
\tilde N^{\dagger}, \tilde S_2, \tilde S^{\dagger}_2, \tilde S_1,
\tilde S^{\dagger}_1)^T$. In general, the $\tilde S_1$'s are
decoupled and have no interactions with the SM particles. In
Ref.\cite{Khalil:2011tb}, it was shown that the lightest
right-handed sneutrino in this model is a viable candidate for
cold dark matter. Also, it was emphasized \cite{Elsayed:2011de}
that the one-loop radiative corrections due to right-handed
(s)neutrinos to the mass of the lightest Higgs boson is
significantly large and can push the upper bound imposed on its
mass to around 200 GeV.

Here we show that right-handed sneutrinos $S_{1,2}$ can be the
inflaton fields and provide a mechanism for chaotic inflation. For
this propose, a gauge singlet superfield, $X$, is introduced (can
be a part of a hidden sector). The
superpotential part, which is relevant for inflation, is given by
%
\begin{figure}[t]
\includegraphics[height=6.5cm,width=4.cm,angle=-90]{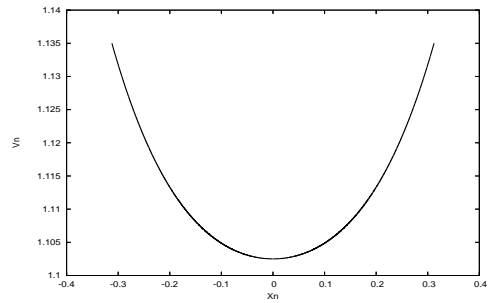}
\vspace{0.25cm} 
\caption{Shape of the Potential ($V_{\rm n} = V/(2m^3M_*)$) along $X$ 
direction ($X$ scaled as $X_{\rm n} = X/(mM_*)^{1/2}$) for $\sigma 
= 1.05 \sigma_c$.} 
\label{plot1}
\end{figure}
%
\begin{equation}
\delta W  =  S_1 S_2 \left(\frac{X^2}{M_*} - m\right),
\label{deltaW}
\end{equation}
where $m$ is a mass parameter (can be large enough) in the model
which would be responsible for providing the vacuum energy during
inflation and $M_*$ is a superheavy scale (can be thought of as
some kind of messenger mass between the hidden and $B-L$ sector).
Note that the renormalizable coupling $X S_1 S_2$ is not allowed
by  $\mathbb{Z}_2$ symmetry, under which $X$ is odd and all other
superfields are even. $X$ has $R$ charge zero. Such a
renormalizable coupling between the $B-L$ sector and singlet $X$
superfield would not lead to a successful model of inflation. As
we have discussed before,  our aim is not to break the $B-L$
symmetry at the end of inflation, since $B-L$ needs to be broken
at a much lower scale $\sim$ TeV.

The corresponding scalar potential can be written as 
{\small 
\begin{equation} V = \left\vert \frac{X^2}{M_*} - m \right\vert^2
\left( |S_1|^2 + |S_2|^2\right) + 4|S_1 S_2|^2 \frac{|X|^2}{M^2_*}
+ {\rm {D-terms}},
\end{equation}}
\noindent where the scalar component of the superfields are denoted by the same symbols as the
corresponding superfields. Note that choosing the direction $|S_1| = |S_2|$ would lead
to vanishing of the D-terms relevant for the discussion of the inflation sector
fields. Specifying this direction as $S = |S_1| = |S_2|$ and choosing the real part
of it as $S = \sigma/\sqrt{2}$,
the potential takes the form,
\begin{equation}
V = \left( \frac{X^2}{M_*} - m\right)^2  \sigma^2 + \sigma^4
\frac{X^2}{M^2_*},
\end{equation}
where the phase of $X$ is taken to be zero for simplicity.
For nonzero $\sigma$, the potential above have extrema along,
\begin{equation}
    X = 0 {\rm {~~or~~}} 2X^2 + \sigma^2 = 2m M_*,
\end{equation}
directions. We note that for $\sigma > \sigma_c = (2mM_*)^{1/2}$,
$X = 0$ corresponds to a local minimum. At this minimum, the
potential $V$ takes the form $V_{\rm inf} =  m^2 \sigma^2$, which can
provide inflation ($\sigma$ is our inflaton). The change in the shape of the 
potential (obtained from $\delta W$) along $X$ direction can be understood from 
Fig.1 and 2 for different values of $S$, above and below $S_c$ respectively.

It is interesting to note that $V_{\rm inf}$ is similar to the potential responsible for chaotic 
inflationary picture and therefore the inflation can be realized for large value of $\sigma$ 
above the reduced Planck scale: $M_P = 2.4 \times 10^{18}$ GeV.  However in our case, 
the dynamics of having this potential is different compared to the chaotic inflation and there 
are more fields involved here. As we have mentioned at the beginning, the construction of 
the inverse seesaw mechanism should not be destabilized by the additional large mass parameter 
$m$ introduced in $\delta W$ ($m \sim 10^{13}$ GeV as we will find soon). This can indeed be 
achieved if $X$ can acquire a vev ($\langle X \rangle$) at the end of inflation. As the expected contribution to the neutrino
mass matrix from $\delta W$ is proportional to $(X^2 - mM_*)$,
a vev of $X, \langle X \rangle = (mM_*)^{1/2}$, at the end of inflation would 
ensure the stability of the neutrino mass through the inverse seesaw. 
As we proceed further, we will find how $X$ can acquire such a vev 
at the end of inflation. So the light neutrino mass is essentially independent 
of the mass parameter involved in the inflationary potential as opposed to the case 
with sneutrino inflation \cite{Murayama:1992ua}. This also explains why a single term 
in the superpotential $m S_1S_2$ which can provide a successful chaotic sneutrino inflation as can not be accommodated alone while working with inverse seesaw mechanism.   

\begin{figure}[t]
\includegraphics[height=6.5cm,width=4.0cm,angle=-90]{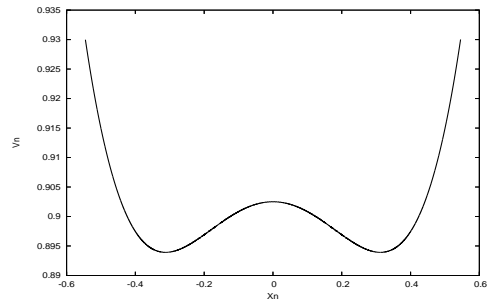}
\vspace{0.25cm} 
\caption{Shape of the Potential ($V_{\rm n} = V/(2m^3M_*)$) along $X$ direction 
($X$ scaled as $X_{\rm n} = X/(mM_*)^{1/2}$) for $\sigma = 0.95 \sigma_c$.} 
\label{plot2}
\end{figure}
%


At the end of inflation, the field $\sigma$ rolls down towards the global minimum 
at $\sigma = 0$. As soon as $\sigma$ is below $\sigma_c$, $X =0$ becomes a local
maximum and a new valley for the $X$ field appears along $X^2
= mM_* - \sigma^2 / 2$ direction, which would be the new local minima.
Therefore $X$ is trapped along that direction only and will stay at this valley 
by continuously adjusting itself in this direction with the change of $\sigma$. It is 
therefore expected to make a smooth transition to $\langle X \rangle = (mM_*)^{1/2}$ at the 
very end when $\sigma$ becomes very small and subsequently goes to zero. 
However we notice that the global minimum lies at $\sigma =0$, thereby making 
$X$ arbitrary. This arbitrariness is actually corrected once we assume a contribution 
from a hidden sector potential carrying interaction (suppressed though) with the 
$X$ field. We will come back on this issue while discussing the post inflationary
phase near the end of this article. With this new contribution appearing from hidden 
sector (not playing any role in inflation), $\langle X \rangle$ will remain same as $(mM_*)^{1/2}$, 
which is instrumental for keeping light neutrino mass matrix uninterrupted.  

We can now discuss the inflationary parameters \cite{review} in connection with
experimental observations. During inflation, $X$ is
at zero and the potential is dominated by $V_{\rm inf} = m^2
\sigma^2$. The slow roll parameters are given by %
\bea %
\epsilon \!=\! \frac{m^2_{Pl}}{16 \pi} \left (
\frac{V'_{\rm inf}}{V_{\rm inf}}\right)^2 \!\approx \!\frac{m^2_{Pl}}{4 \pi \sigma^2},
~~~ \eta  \!=\! \frac{m^2_{Pl}}{8 \pi} \left (
\frac{V''_{\rm inf}}{V_{\rm inf}}\right) \!\approx\! \frac{m^2_{Pl}}{4 \pi \sigma^2},~ %
\label{param} %
\eea %
where $m_{Pl}$ is the Planck scale $\sim 1.22 \times 10^{19}$ GeV. Note 
that the slow roll parameters are independent of the mass scale $m$.   
The number of e-foldings is provided through the relation, %
\bea %
N \! =\! \frac{8\pi}{m^2_{Pl}} \int_{\sigma_e}^{\sigma_0} \frac{V_{\rm inf} d\sigma}{V'_{\rm inf}} \! \approx \! \frac{2 \pi}{m^2_{Pl}} \left ( 
\sigma^2_0 - \sigma^2_e \right ), ~ %
\label{param} %
\label{N} %
\eea %
where the subscript ${\footnotesize e}$ corresponds to the values at 
the end of inflation and ${\footnotesize 0}$ indicates the values 
when comoving scales comparable to our present horizon crossed outside the inflationary horizon ({\it i.e.} when 
the comoving wavenumber $k_0 = 0.002$ Mpc$^{-1}$). Hence the slow roll parameter at that time can be expressed as 
$\epsilon_0 =$ 1/(1 + 2$N$), 
where we have used the fact that $\epsilon_e \sim 1$. So with $N \simeq 60$, $\epsilon_0 \simeq 0.008$ at that time.  
The amplitude of curvature perturbation $\Delta_R$ is given by,
\begin{equation}
\Delta_R = \frac{8}{m^3_{Pl}} \sqrt{\frac{2 \pi}{3}} \frac{V_{\rm inf}^{3/2}}{|V'_{\rm inf}|}.
\label{D}
\end{equation}
According to WMAP seven years data\cite{Komatsu:2010fb},
$\Delta_R \sim 4.93 \times 10^{-5}$ corresponding to
$k_0$. Therefore from eq.(\ref{N}) and eq.(\ref{D}), we find
that $m \sim 1.3 \times 10^{13}$ GeV. Now using the expression for the
spectral index for scalar density perturbation, $n_s = 1 - 6\epsilon + 2\eta$,
we find $n_s \sim  0.97$, which is in good agreement with recent WMAP data \cite{Komatsu:2010fb}. The 
prediction for the ratio of scalar and tensor perturbations $(r)$ at the quadrupole scale is found to be $r \sim 0.13$ which is consistent with the
WMAP findings \cite{Komatsu:2010fb}.

We are now in a position to discuss the post inflationary phase
and reheating. We have seen that for $\sigma < \sigma_c$,
$X$ is trapped in the valley, $X^2 = mM_* - \sigma^2/2$ and 
when $\sigma$ approaches zero, the field $X$ should smoothly enter 
into the minimum at $X = (mM_*)^{1/2}$. However $S = 0$
(or $\sigma = 0$) makes X arbitrary as at this stage $X$ field 
is massless. It results to a possibility that $X$ can be 
deviated from its would be minimum at $(mM_*)^{1/2}$, which is unwanted 
for keeping the neutrino mass in the right range. In order to cure 
this problem, we introduce a term in the scalar potential, originated from an  interplay between a hidden sector and $X$, 
\begin{equation} V_h (X, Y) = h^2 |Y|^2 \left( X - M \right)^2, 
\label{Vh}
\end{equation}
where $Y$ is a superfield in the hidden sector, $M$ is a mass
scale related by $M^2 = mM_*$ and $h$ is the coupling constant. 
This choice of $M$ ensures that $X$ minimum is not shifted anymore 
from $X = (mM_*)^{1/2}$ provided $\langle Y \rangle \neq 0$. This relation 
may not be completely unnatural as the other mass parameter involved 
in the scenario, $M_*$, is not restricted from inflation. At this point we do 
not attempt to construct any specific model for the hidden sector so as to 
realize $V_h$ directly. We 
assume that $Y$ get this vev when $\sigma < \sigma_c$,  
so that the inflationary phase is not perturbed by the introduction 
of $V_h$ (during inflation $Y$ is expected to be at the origin). 

Therefore in this picture, during early stage of inflation and
thereafter, both $S$ and $X$ have field dependent masses which are
decreasing as $S$ approaches zero. Following our assumption, at 
some stage (when $X$ is already in the valley of $X^2 = mM_* - \sigma^2/2$), 
$X$ starts to receive an additional contribution towards its mass 
from $V_h$ with $ h \langle Y \rangle = m_X$. This massive $X$ will oscillate around its 
minimum and can therefore decay and reheat the universe. From 
eq.(\ref{deltaW}), we note that a decay of $X$ 
to fermionic components of $S_1, S_2$ is possible and can be estimated as 
$\Gamma_1 \approx (h/\pi) (M/M_*)^2 \langle Y \rangle$. For a choice of 
$M_* \sim 10^{16}$ GeV (close to the GUT scale) with $h \sim O(10^{-2})$, 
the reheat temperature, $T_R \simeq \frac{1}{7}
{\sqrt{\Gamma M_P}}$, is found to be $\lesssim 10^9$ GeV provided 
$\langle Y \rangle \lesssim 10^{7}$ GeV. 

From eq.(\ref{Vh}), it can be found that the $X$ field may also decay 
into $Y$ and thereby reheating could be into the hidden sector only, 
provided this decay mode is the most efficient. However in order to 
achieve a successful reheating of the observable universe, we can 
assume that the mass of the scalar component of the $Y$ field 
($m_Y$) is bigger than $m_X$. So this decay and hence the reheating 
into the hidden sector only can be prevented. We can safely assume 
this ($m_Y > m_X$) as $m_Y$ would depend on the constructional details 
of the hidden sector itself. There could be another way out if the $Y$ 
field is part of the supersymmetry breaking sector. In this case 
through the presence of the mediator of the SUSY breaking connecting the hidden as well as the observable sector, 
the universe can be finally reheated as shown in \cite{arun}. 
Another possibility is the preheating \cite{preheating} of the universe 
after inflation through the parametric resonance. However this would 
involve several other fields in our setup and requires detailed study. 
At this point, the detailed discussion of reheating is beyond the scope 
of this article.

In conclusion, we have constructed a model for sneutrino inflation
embedded in SUSY $B-L$ extension of the SM along with an additional 
singlet field, where the light neutrino mass is generated through inverse 
seesaw mechanism. 
At an early stage of the universe, these sneutrino fields are having 
very large values and inflation is initiated. $B-L$ symmetry is
therefore broken at that time. However when inflation is over, the inflaton 
rolls down to zero and additional singlet field receives a nonzero vacuum 
expectation value at the very end, thereby the $B-L$ symmetry is restored. 
This $B-L$ symmetry will be broken radiatively at a TeV scale and the 
electroweak symmetry breaking will take place subsequently. Then the
light neutrino mass is generated at the expected level. 
It can be noted that the interplay of two fields involved in realizing inflation bears
a similarity with the supersymmetric hybrid inflation models (SHI) \cite{shi}. In 
SHI, the superpotential is linear in the inflaton superfield. So the inflationary 
direction is absolutely flat and a slope needs to be generated using the 
Coleman-Weinberg correction. On the contrary, there is a nonvanishing potential existing for 
the inflaton in our case along which it rolls down naturally towards its minimum at zero.

\section*{Acknowledgments}
The work of S.K. was partially supported by the Science
and Technology Development Fund (STDF) project 1855 and the ICTP
AC 80. A.S. is grateful to the Abdus Salam ICTP for their 
hospitality during a visit when this work was initiated. A.S. is 
partially supported by the Start-Up grant from IIT, Guwahati.

\end{document}